\documentclass[10pt,preprint2]{aastex}

\newcommand\plotonerot[1]{%
 \centering 
 \leavevmode 
 \includegraphics[angle=90,width=\columnwidth]{#1}%
}%

\def\ion#1#2{\mbox{\rm #1\sc #2}}

\def\HII{{\ion{H}{ii}}}

\def\dim#1{\mbox{\,#1}}
\def\figdir{.}

\begin{document}


\title{Lyman Break Galaxies and Reionization of the Universe}
\author{A.\ Gayler Harford and Nickolay Y.\ Gnedin}
\affil{Center for Astrophysics and Space Astronomy, 
University of Colorado, Boulder, CO 80309}
\email{gharford@casa.colorado.edu, gnedin@casa.colorado.edu}

\begin{abstract}
   We present the star formation histories, luminosities,
colors, mass to light ratios, and halo masses of "galaxies"
formed in a simulation of cosmological reionization.  We
compare these galaxies with Lyman Break Galaxies
observed at high redshift.  While the simulation is
severely limited by the small box size, the simulated
galaxies do appear as fainter cousins of the observed ones.
They have similar colors, and both simulated and observed
galaxies can be fitted with the same luminosity function.
This is significant to the process of reionization since
in the simulation the stars alone are responsible for
reionization at a redshift consistent with that of the
Gunn-Peterson trough seen in the absorption spectra of
QSO's.  The brightest galaxies at $z = 4$ are indeed quite
young in accord with observational studies.  But these 
brightest galaxies are free riders - they contributed 
only about 25\% to the reionization of the universe at 
$z > 6$.  Instead, the bulk of the work was done by dimmer
galaxies, those that fall within the $28 < I < 30$ magnitude
range at $z\sim4$.  These dimmer galaxies are not necessarily
less massive than the brightest ones.
\end{abstract}

\keywords{cosmology: theory - cosmology: large-scale structure of universe -
galaxies: evolution - galaxies: formation - galaxies: high-redshift - 
galaxies: starburst - galaxies: stellar content - infrared: galaxies}

\section{Introduction}

   The most dramatic global transition in the recent 
history of the universe is the reionization of the
intergalactic medium.  The recent identification of
the predicted Gunn-Peterson trough in the absorption
spectra of several QSO's has placed this event at
about a redshift of 6 (Djorgovski et al.\ 2002; Becker et al.\ 2001).  
   
   Ionizing radiation from newly
formed stars has been suggested as the most likely
source of this radiation.  Bright, optically selected
QSO's appear to be too sparse during the relevant 
period (Fan et al.\ 2001).  Low brightness AGN's
remain a possibility, but no other strong candidates
have emerged.
   
   It is natural to look for evidence for these
ionizing stars in the accumulating observational
data on high redshift ($z\la4$) galaxies (Madau et al.\ 1996; 
Madau, Pozzetti, \& Dickinson 1998; Steidel et al.\ 1999; 
Ouchi et al.\ 2001; Ferguson, Dickinson, \&
Papovich 2002; Hu et al.\ 2002; Malhotra \& Rhoads 2002;
Lehnert \& Bremer 2003; Kodaira et al.\ 2003; Iwata et al.\ 2003). 
Some studies on Lyman Break Galaxies (LBG)
have questioned whether enough stars could have been 
formed to accomplish reionization at so early a time
(Ferguson, Dickinson, \& Papovich 2002; Dickinson et al.\ 2003).
   
   In the popular Lambda-CDM cosmological model
the seeds of galaxy formation lie in minute
density fluctuations, which manifest themselves
as temperature fluctuations in the Cosmic Microwave
Background (CMB).  As the fluctuations continue
to grow into dark matter halos, they trap gas
within their potential wells.  That part of the
gas that is able to radiate its energy efficiently
eventually condenses into dense molecular clouds,
that give birth to stars.
   
   Reionization has been studied within this
paradigm both by semi-analytical methods (Giroux \&
Shapiro 1996; Ciardi \& Ferrara 1997; Haiman \& Loeb
1997, 1998; Madau et al.\ 1999; Valageas \& Silk 1999;
Chiu \& Ostriker 2000; Miralda-Escude, Haehnelt \&
Rees 2000; Barkana \& Loeb 2000; Oh et al.\ 2001; 
Wyithe \& Loeb 2003) and by numerical simulations 
(Ostriker \& Gnedin 1997; Gnedin \& Ostriker 1998; 
Gnedin 2000; Nakamoto, Umemura, \& Susa 2001; 
Razoumov et al.\ 2002).  The former
necessarily involve broad simplifying assumptions
and, therefore, have limited predictive power.
The latter, although more realistic and fully
self-consistent, are limited by the dynamical range
that can be accommodated by existing numerical
techniques on modern supercomputers.  The two
approaches are complementary and will have to 
remain so until the next generation of cosmological
numerical codes become available on faster
computers.
   
   For this paper we use a
simulation that was reported in Gnedin (2000).
The simulation includes dark matter, gas, star
formation, chemistry and ionization balance in
the primordial plasma, and an approximate
implementation of 3D radiative transfer.  All
galaxies with total masses in excess of $3\times10^8\dim{M}_\sun$
are resolved both spatially and by mass. The
simulation was continued until $z=4.0$.

   The overall picture for the evolution of the
early universe that has emerged from this simulation
is as follows:

      Star formation first begins in earnest at about $200\dim{Myr}$
      ($z\sim20$).  The overall star formation rate
      increases to a maximum at about $1\dim{Gyr}$
      ($z\sim5.5$) and decreases slightly thereafter.

      Reionization of the intergalactic medium
      occurs over a prolonged period of time from
      about $350\dim{Myr}$ ($z\sim12$) to about 
      $900\dim{Myr}$ ($z\sim6$),
      with expanding \HII\ regions in the low density
      IGM merging at $z\sim7$.  This latter
      moment of rapid transition is often
      identified as the "moment of reionization".
      This moment approximately corresponds to the mass (volume) averaged
      neutral fraction in the low density IGM falling below 1 (0.1) percent 
      respectively. 

      As a result of reionization, the gas content
      of halos less massive than about $3\times10^9\dim{M}_\sun$
      solar masses is drastically reduced.

   Since this simulation does not include AGN's, it
provides a detailed example of how stars alone might
be sufficient to accomplish reionization.  The 
addition of AGN's to this scenario would presumably
lead to even more ionizing photons, which could take
up the slack in the event that the amount of star
formation in the simulation were overestimated.

   The simulation provides us with a detailed star
formation history for each galaxy and thus allows
us to make detailed photometric predictions. In this
paper we present the star formation histories,
luminosities, colors, mass to light ratios, and halo
masses of the simulated galaxies.  

   Our simulation, while state of the art, is 
only barely sufficient for our purpose, because the 
computational box is so small.  This inherent 
limitation of any simulation limits the number of 
bright objects that can be modeled, while the
observations approach the galaxy hierarchy 
from the other end, detecting bright galaxies
and missing the dim ones.  In this sense we
can think of observations and simulations as
moving toward each other along the galaxy luminosity
curve.  Although there is not much overlap yet,
the junction point has at last been reached, as we
show below.

   We argue that to the extent comparisons can be 
made, our predictions are consistent with observations 
of Lyman Break Galaxies.  The varied star formation 
histories that we see provide the key to understanding 
why the observations appear to argue that LBG galaxies 
are too young to have reionized the universe.

\section{Method}

\subsection{Simulation}

The specific simulation we use in this paper is fully described in
Gnedin (2000). The simulation was performed using the Softened Lagrangian
Hydrodynamics (SLH) code, which includes dark matter, gas, star formation,
chemistry and ionization balance in the primordial plasma, and
 3D radiative
transfer (in an approximate implementation). All these physical ingredients
are required to properly model the process of cosmological reionization. 

The simulation of a representative CDM+$\Lambda$ cosmological 
model\footnote{With the following cosmological parameters: $\Omega_0=0.3$,
$\Omega_\Lambda=0.7$, 
$h=0.7$, $\Omega_b=0.04$, $n=1$, $\sigma_8=0.91$, where the amplitude and
the slope of the primordial spectrum are fixed by the COBE and cluster-scale
normalizations.}
was performed in a comoving box with the size of $4h^{-1}\dim{Mpc}$ with the
total mass resolution of $4\times10^6\dim{M}_\odot$ and the comoving
spatial resolution of $1h^{-1}\dim{kpc}$. This resolution is a reasonable
compromise between the need to have the box large enough to accommodate
several \HII\ regions and the need to have spatial resolution high enough 
to resolve sources of ionizing radiation. Convergence
studies presented in Gnedin (2000) demonstrate that, while still limited, 
the resolution of this simulation is sufficient to model reionization at a
semi-qualitative level (with precision of the order of 50\% or so).

The simulation was stopped at $z=4$ because at this time the rms density 
fluctuation in the computational box is about 0.25, and at 
later times the box
ceases to be a representative region of the universe even for the dark 
matter.

The simulation was designed in such a way as to approximately 
reproduce the measured star formation rate density at $z=4$ of 
about $0.1\dim{M}_\odot/\dim{yr}/\dim{Mpc}^3$ (Steidel et al.\ 1999).

Our simulation assumes that all cosmological reionization occurs via
radiation from {\em stellar} sources, with star formation parameterized
by the phenomenological Schmidt law as discussed in Gnedin (2000a). An
alternative and complementary scenario would be one in which the bulk of
reionization is produced by Active Galactic Nuclei (AGN). In light of recent
high-redshift quasar counts (e.g., Fan et al.\ 2001), a scenario in which
the universe is reionized by bright optically selected QSOs seems
unlikely in any event. There remains however the possibility that 
low brightness AGNs contributed significantly (if not dominantly) to the
reionization of the universe. 

For the purpose of this paper, however, this simulation is quite suitable
for the following reason. Our main goal in this paper is to investigate
whether galaxies are capable of reionizing the universe and still appear as
young as they do at $z\sim4$. Thus, by ignoring the AGN component, we
consider the most extreme case, since we require galaxies to carry all the
burden of reionizing the universe by themselves. One would then expect that
if galaxies are capable of reionizing the universe {\em and\/} appear as
young as they do at $z\sim4$, then it will be even easier for the models
to agree with observations if some ionization is done by AGNs.

We identify galaxies in the simulation with gravitationally bound objects,
which are selected with the DENMAX algorithm of Bertschinger \& Gelb
(1991).

\subsection{Population Synthesis}

	Population synthesis is carried out using the 
Starburst99 package (Leitherer et al.\ 1999).  The version we use
included the July 2002 update, which revised the UV spectra 
for O and B stars.  These spectra,
based on the recent work of Smith, Norris, \& Crawther (2002),  
have decreased 
ionizing radiation during the early
stages of star formation.    These changes affect not only 
the amount of Lyman alpha 
continuum, but also the amount of Lyman alpha that can be 
produced if this UV is absorbed by local gas.  We choose the 
instantaneous burst option.  This is appropriate because 
we apply spectral synthesis to individual stellar particles formed
in the simulation, which sample the stellar distribution function at
discrete time steps. 

   We choose the lowest metallicity option (5\% solar) available in
Starburst99, 
since the simulated galaxies have comparable metallicities.
Starburst99 maintains
same metallicity for each subsequent
generation of stars.  This simplification should be
sufficient for times less than about $1\dim{Gyr}$ (Leitherer et al.\ 1999).

	Ultra-violet absorption by the IGM material along the 
line of sight (i.e.\ by the Lyman alpha forest) is taken 
into account using the method of Madau (1995).

	Reddening corrections are made by the method of 
Calzetti (1997), including the recent calibration data of 
Leitherer et al.\ (2002) 
We find it sufficient to consider a range of reddening
corresponding to a color excess of $E(B-V) = 0 - 0.3$.  Higher amounts 
of reddening produce colors inconsistent with nearly all 
the observed galaxies.  This range is in line with the
estimate of Shapely et al.\ (2001) 
of an average color excess of $E(B-V)=0.15$ 
for a sample of Lyman break galaxies at a redshift of about 3.  
This range is also generally consistent with the results
of others for high redshift galaxies.

We also allow for the partial escape of ionizing photons from the
modeled galaxies by reducing the flux above $13.6\dim{eV}$ by a factor
$f_{\rm ESC}$, and converting 2/3 of the removed photons into
Lyman-alpha emission following the recipe of Malhotra \& Rhoads
(2002). The escape fraction $f_{\rm ESC}$ becomes another parameter
in our modeling. We consider three values for this parameter:
0.1, 0.5, and 1.0.

   The $AB$ magnitude system is used throughout. We adopt the same
cosmology as used in the simulation
(flat concordance cosmology with $H_0=70\dim{km/s}/\dim{Mpc}$.

   For comparison purposes the $G-I$ colors for observed 
galaxies over a range of redshifts were corrected to a 
redshift of 4.0.  This was done as follows.  For each of a 
range of $z$ values, including 4.0, a series of 199 model galaxies 
of different ages
was constructed at $5\dim{Myr}$ intervals, each having a single 
burst of star formation. For each observed galaxy the set 
of models at the nearest $z$ value was selected.  The two 
bracketing models at this redshift plus the two corresponding 
models at redshift 4.0 were used to obtain a corrected color
using a linear interpolation.  This method was feasible 
because the relationship between $G-I$ and age
is, with few exceptions at these intervals, monotonic.  A 
similar correction for ${\cal R}-I$ was not possible.

\begin{figure}[t]
\plotonerot{\figdir/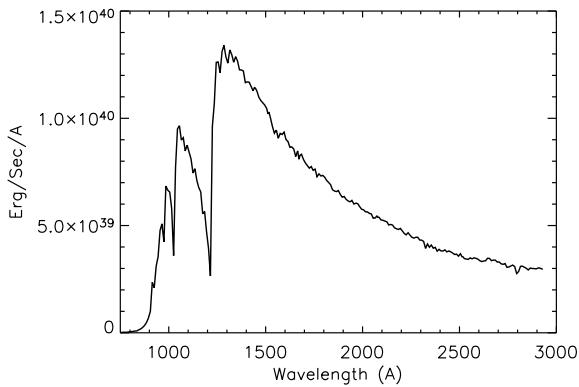}
\caption{\label{fig0}
Calculated spectrum of a typical bright simulated galaxy at redshift 4.0.
Ultra-violet absorption by the Lyman alpha forest has been taken
into account. An escape fraction of 1.0 was used with no
reddening correction.
}
\end{figure}
Figure \ref{fig0} shows an illustrative spectrum of a typical bright galaxy
from our simulation. Sharp absorption edges due to intervening
Lyman-$\alpha$ and Lyman-$\beta$ forests are clearly visible.

\section{Results}

\begin{figure}[t]
\plotonerot{\figdir/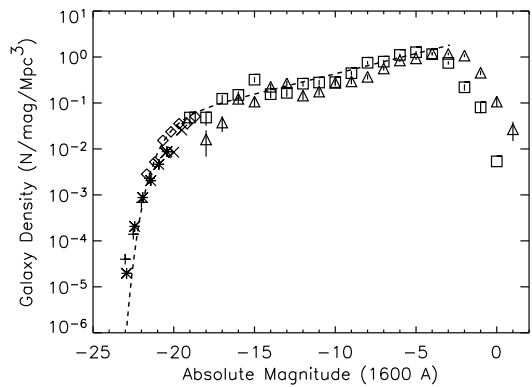}
\caption{\label{fig1}Luminosity Function at rest frame 
1600 A  for simulated galaxies 
at $z=4.0$ in the $I$ band. Squares show the unreddened luminosity
function,
while triangles give the luminosity function with 
$E(B-V)=0.15$ reddening included. Vertical error-bars are 1-sigma Poisson errors. Diamonds
show the observed luminosity function of HDF galaxies from Madau et al.\ 
(1996), while crosses and asterisks mark the observed LBG luminosity
functions at $z\sim4$ and $z\sim3$ respectively 
from Steidel et al.\ (1999).
The dashed line is our Schechter function fit to the combined
luminosity function.}
\end{figure}
	Figure \ref{fig1} shows a computed luminosity function for  
the simulated galaxies at $z=4.0$.  As a measure of luminosity 
we use the $I$ band $AB$ magnitude, which corresponds to a rest frame
wavelength of about 1600 A.  Also shown for comparison is the 
observed luminosity function for Lyman Break Galaxies at 
$z\sim3 - 4$ from a variety of authors (Madau et al.\ 1996; Steidel et al.\ 1999).
Because of the difficulties in observing at these redshifts, 
Steidel et al.\ (1999) have not attempted to make a Schechter fit
at $z=4$,
but instead have observed that the data are well described by 
the Schechter function fitted to Lyman Break Galaxies at 
a redshift of about 3. This fit, however, does not match the 
slope of the luminosity function of simulated galaxies at faint
magnitudes, but the combined luminosity function (simulated plus observed)
can be fit by a single Schechter
function with a shallower slope. The parameters of both fits are listed
in Table 1.

\begin{table}
\caption{Schechter function fit parameters}
\begin{tabular}{ccc}
\tableline
\tableline
Parameter & Steidel et al.\ fit & Our fit\\
\tableline
$\alpha$ & -1.60 & -1.22 \\
$M_*$ & -20.9 & -20.4 \\
$\phi_*\,(\dim{Mpc}^{-3})$ & $1.8\times10^{-2}$ & $5.7\times10^{-2}$ \\
\tableline
\end{tabular}
\end{table}

	Because of the small size of the simulation box, only 
a few of the observed points are expected to overlap the 
simulated ones, and these are all from the Hubble Deep Field
North, which probed fainter magnitudes than did the ground 
based observations of Steidel et al.\ (1999). But it appears
that both the simulation and the observations can be fitted with
the same luminosity
function, implying that the simulation produces galaxies similar to the
real ones. 
The limitation of the small box size directly translates into
the upper limit on galaxy luminosities (there should be at least a few
galaxies 
of luminosity $L$ in the simulation box to determine $\phi(L)$). As one can
see, our simulation has just enough volume to reach the luminosities of HDF
galaxies, and is a factor of 2 too small to have a single galaxy as large
as those found by Steidel et al.\
(1999) in the simulation volume. Thus, we are not able to make
comparisons on a galaxy-by-galaxy basis yet, but future simulation will be
able to model volumes comparable to those covered by observations in only 
a few years. 

Given a Schechter function fit, we can integrate the total luminosity
function to estimate the fraction of starlight in observed and simulated
galaxies. We find that the observed galaxies contain about 2/3 of the total
luminosity density, while the simulated ones account for about 1/3. 
   The contribution of observed galaxies to reionization
is expected to be somewhat smaller however (Gnedin 2000b). 
This is because these galaxies tend to be clustered
(Steidel et al.\ 1999; Adelberger et al.\ 1998), and
most of the time several of them sit inside the same
\HII\ region.  When an \HII\ region grows to be larger
than the mean free path of ionizing photons, these
photons are wasted.

   These arguments imply that the redshift of 
reionization in the simulation is an underestimate,
but not by a large factor because the star formation
rate in the simulation increases rapidly with time
(Gnedin 2000a).
If we assume that the simulation underestimates
the volume filling fraction of the ionized gas at any given time by, say,
50\%, then the
redshift of reionization is underestimated in the simulation
by about $\Delta z\sim1.5$.
As we emphasized above, our simulation is clearly insufficient to give
a quantitatively accurate model of reionization, so this level of precision
is more than acceptable for our purposes.

A remarkable feature of Fig.\ \ref{fig1} 
is flattening of the luminosity function in
the dwarf galaxy range. This effect is the result of the inhibited
accretion of gas onto low mass halos because of photoionization,
sometimes improperly called ``photoevaporation'' (Thoul \& Weinberg 1996;
Quinn, Katz, \& Efstathiou 1996; Weinberg, Hernquist, \& Katz 1997; Navarro
\& Steinmetz 1997; Gnedin 2000b; Chiu, Gnedin, \& Ostriker 2000; Sommerville
2002; Benson et al.\ 2002a,b), and has been reproduced by earlier
simulations (Nagamine 2002).

\begin{figure}[p]
\plotonerot{\figdir/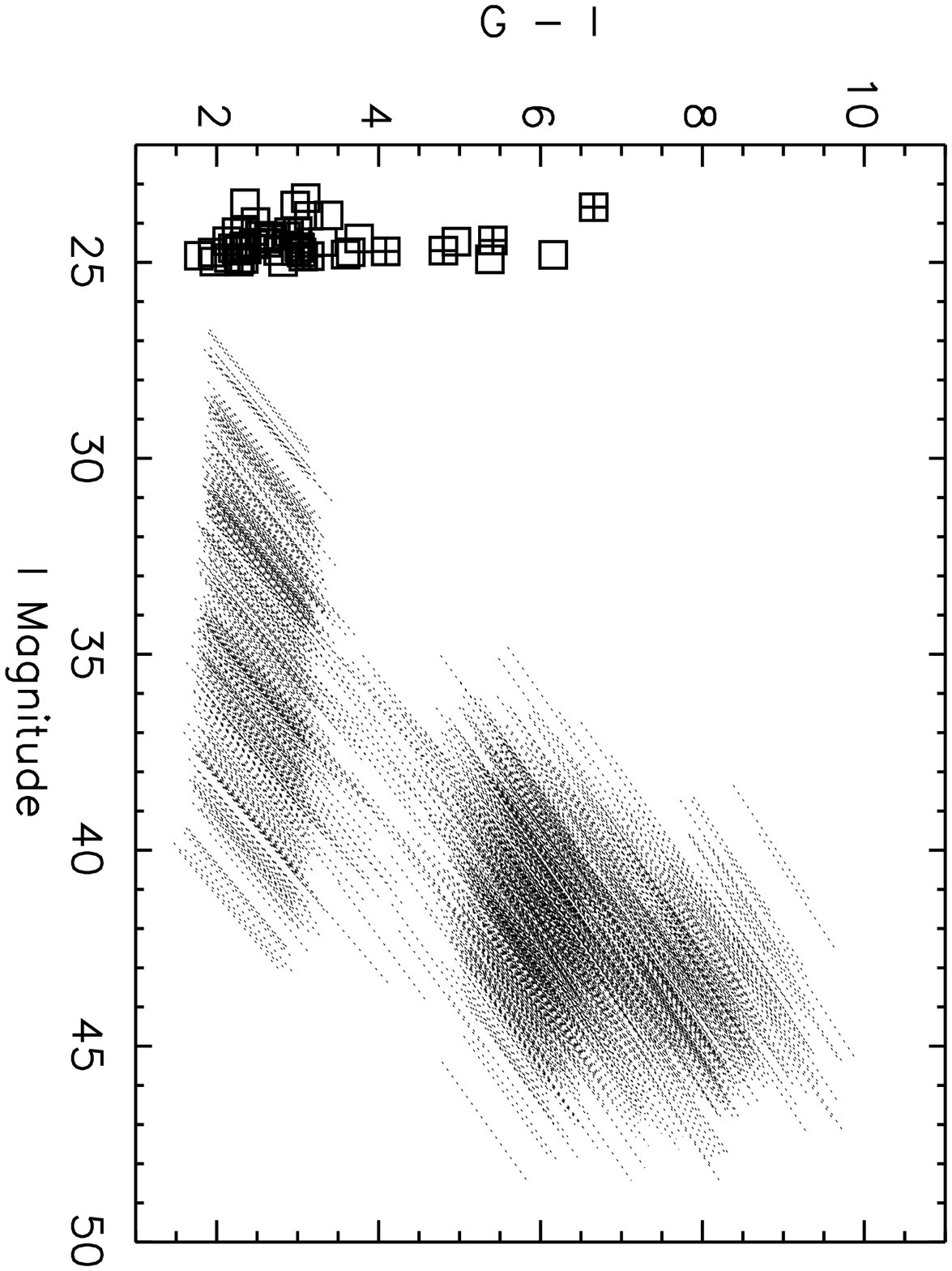}
\plotonerot{\figdir/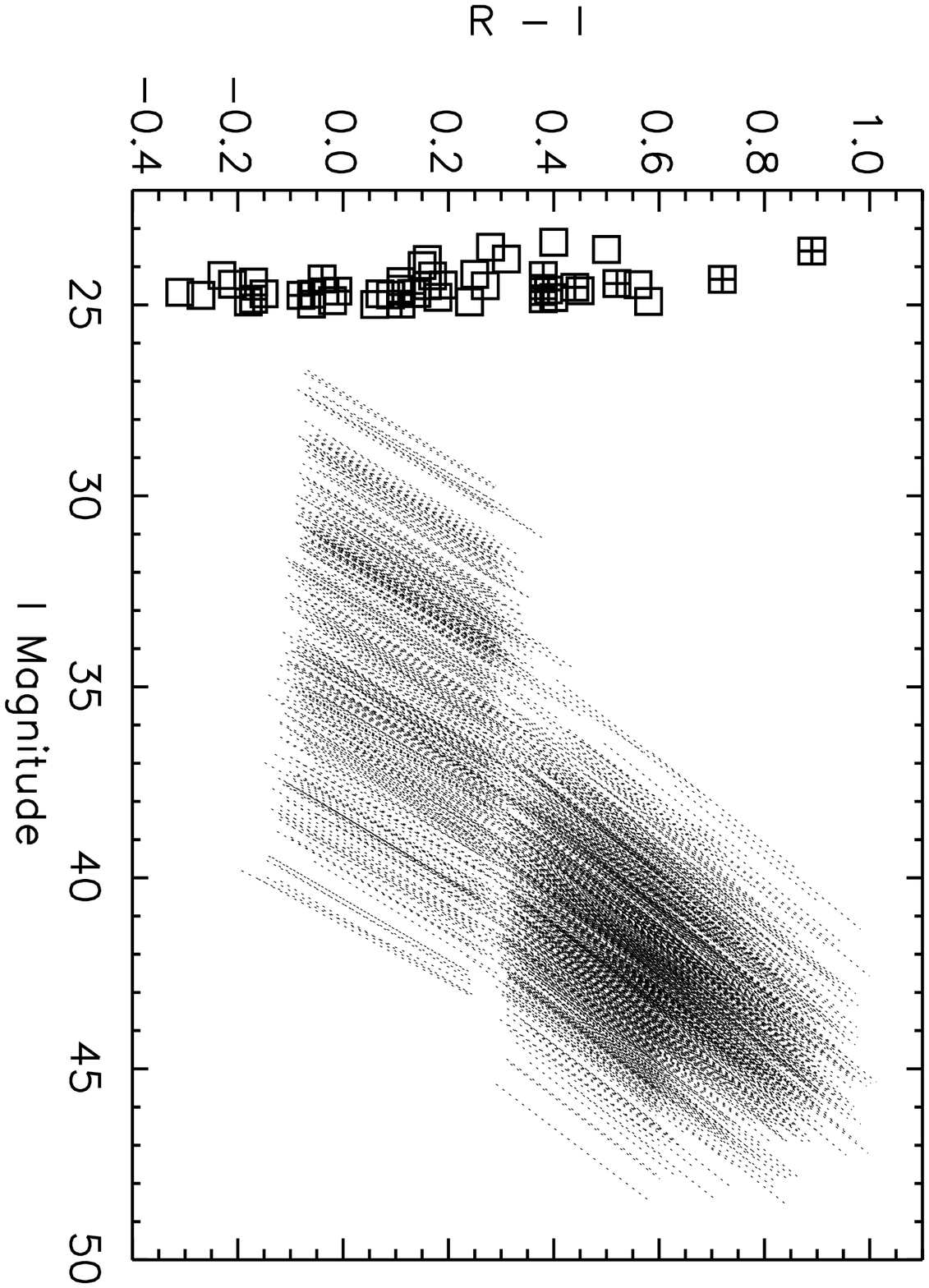}
\caption{\label{fig2}Color-Magnitude Diagrams for Simulated Galaxies at 
$z=4.0$ shown as ({\it a\/}) $G-I$ bv $I$ and ({\it b\/})
${\cal R}-I$ vs $I$.
Each galaxy is represented by a region spanning the range of
of reddening from $E(B-V)=0$ to $E(B-V)=0.3$ and the range of escape
fractions from $f_{\rm ESC}=0.1$ to $f_{\rm ESC}=1.0$.
Squares are observed LBGs from Steidel et al.\ (1999) in the redshift
range from $z=3.5$ to $z=4.8$. A cross within
the square indicates that only a minimum $G$ magnitude could be measured.}
\end{figure}
Figure \ref{fig2} shows Color-Magnitude Diagrams (CMD) for the simulated
and observed galaxies. Because the simulation cannot predict the redenning
correction, we plot each simulated galaxy as a dotted region (which appears
almost like a single line), spanning a
range
in redenning corrections from $E(B-V)=0$ to $E(B-V)=0.3$, consistent with
values found by Steidel et al.\ (1999), and a range of escape fractions
from $f_{\rm ESC}=0.1$ to $f_{\rm ESC}=1.0$.

One can see that simulated galaxies have similar colors to the observed
ones, which suggests that the simulation captures at least some 
of the physics, which is going on in these galaxies.
Let us now imagine a ``super-simulation'', with so large a box that it
can be considered infinite. 
Because the colors of our brightest galaxies do
not change much with magnitude, it is plausible to assume that simulated
galaxies in the magnitude range $23<I<25$ would have similar colors to the
observed ones. If we plausibly assume that the simulated luminosity
function, when extended to higher luminosities, would match the observed one
reasonably well, then the total luminosity density in such a
``super-simulation'' would be a factor of 3 higher than
in our simulation, and, thus, the redshift of reionization would be even
higher than $z=7$ - the redshift of reionization in our simulation - by
about $\Delta z\sim1.5$. Such a
``super-simulation'' would serve as a strong counter example to claims that
LBGs at $z\sim4$ are too young to reionize the universe by $z\sim 6$.

\begin{figure}[p]
\plotonerot{\figdir/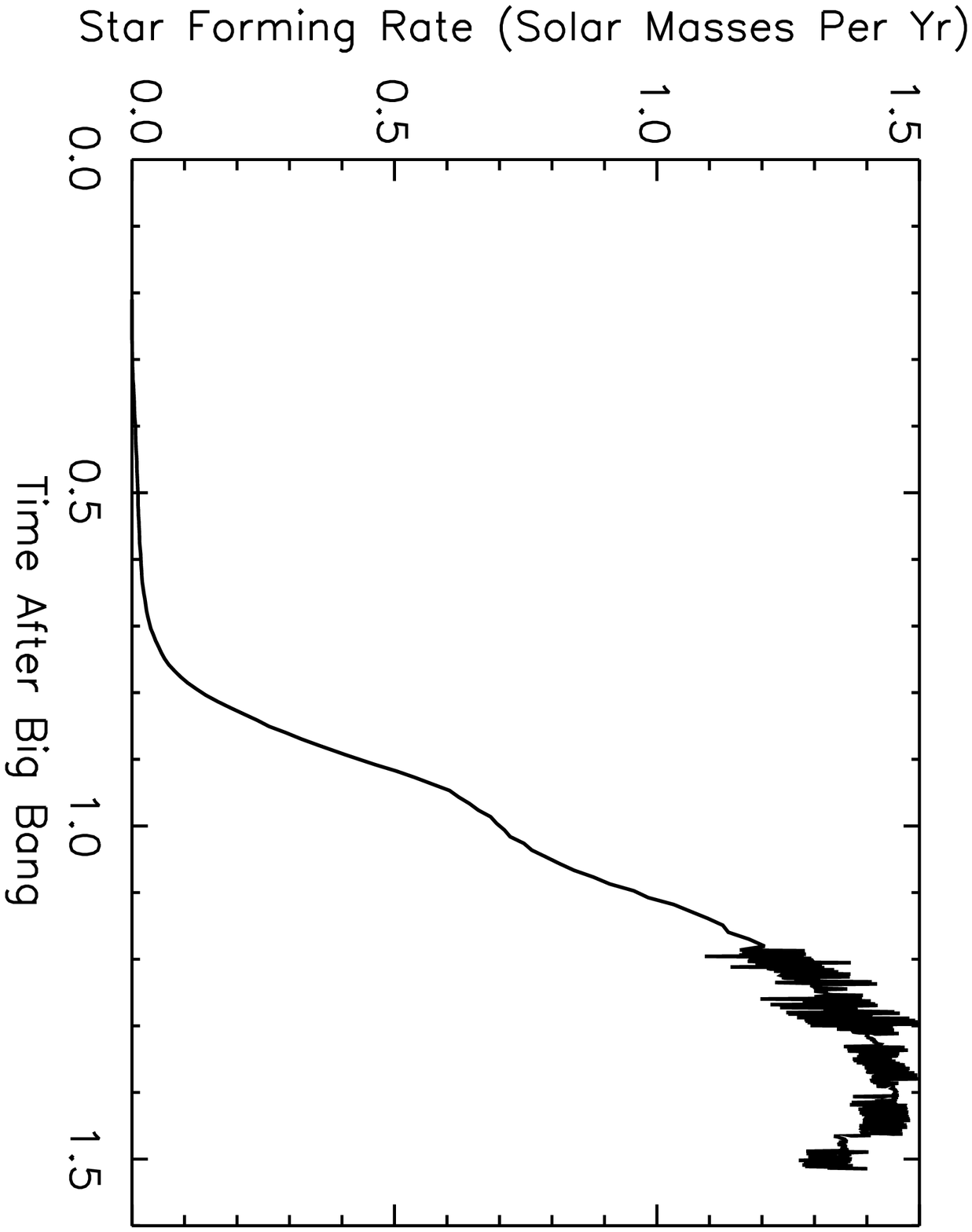}
\plotonerot{\figdir/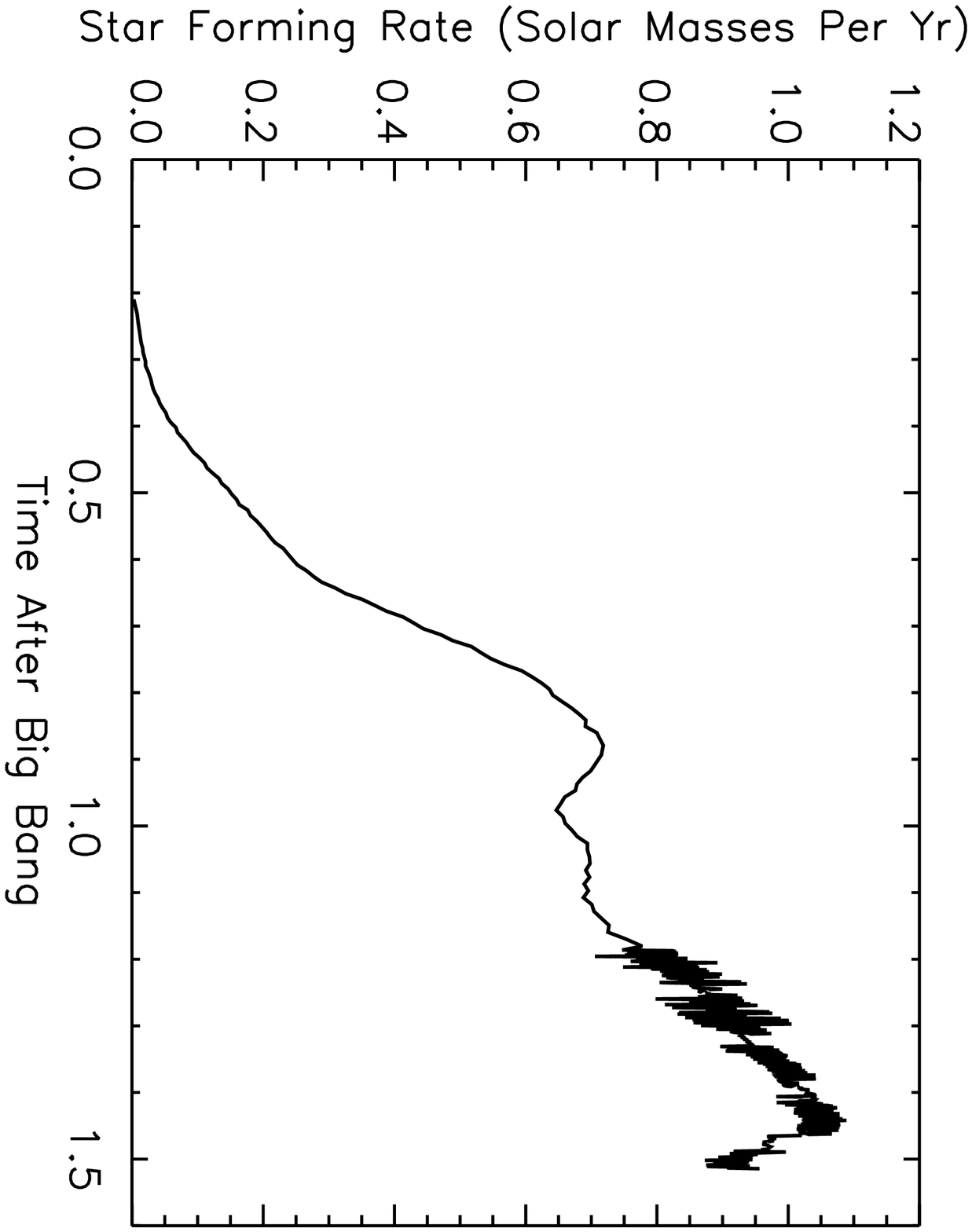}
\plotonerot{\figdir/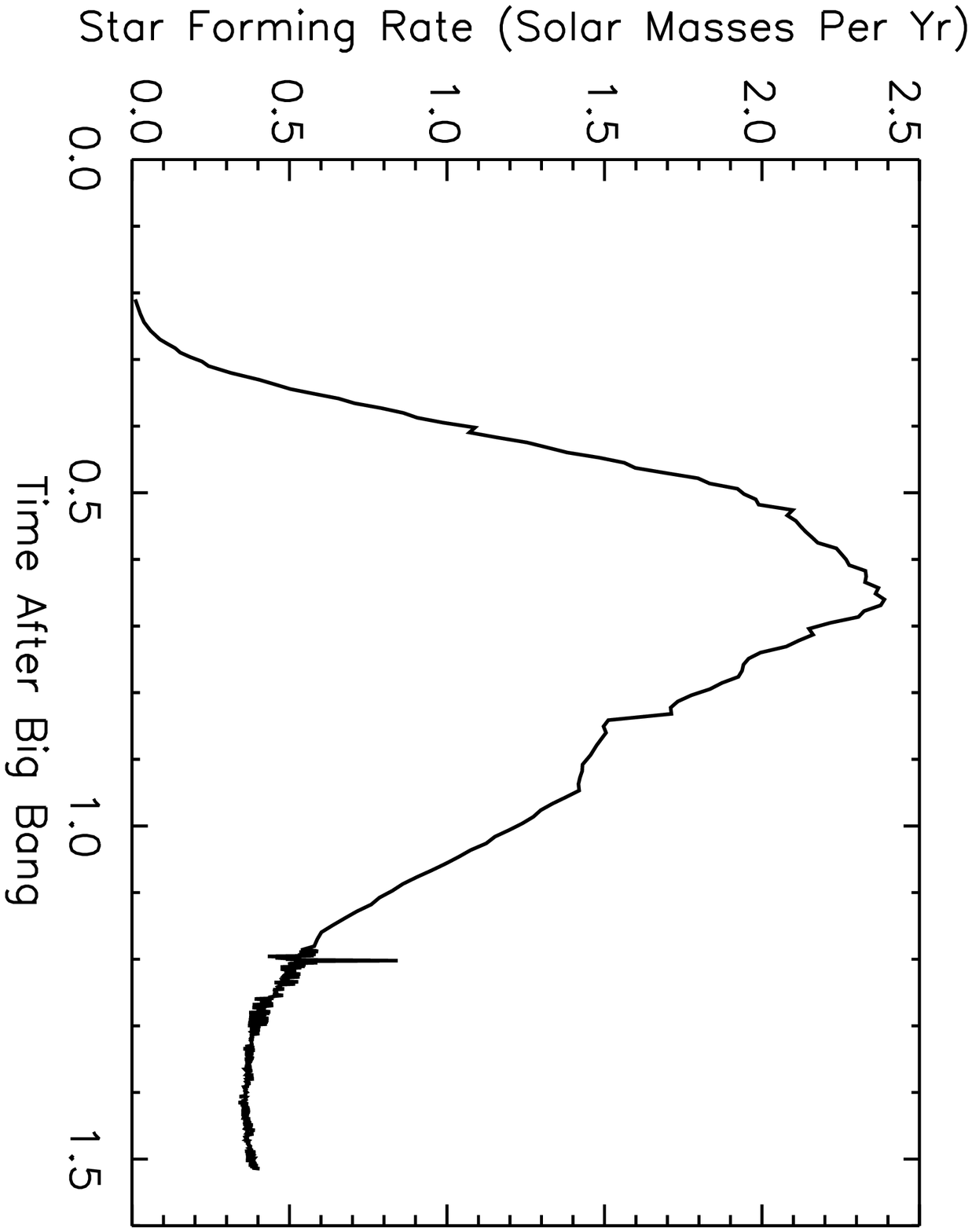}
\caption{\label{fig3}
Three illustrative star formation histories for simulated galaxies with
different luminosities: 
$I=26.9$ ({\it a\/}),
$I=27.2$ ({\it b\/}), 
$I=28.0$ ({\it c\/}).
Despite the very
different histories, all three have very similar $G-I$ and ${\cal R}-I$
colors. 
The temporal sampling of star formation histories is 10 times higher after
$1.2\dim{Gyr}$, showing fine structure variability on a few Myr time scale
due to formation of individual star clusters, which are resolved (by mass)
in the simulation.
}
\end{figure}
Why, then, is observational determination of ages so misleading? 
Figure \ref{fig3} illustrates the diversity of star formation histories in
simulated galaxies. While our brightest galaxies ($I\la27$, such as the
one shown in 
panel {\it a\/}) 
are experiencing a peak of star formation at around $z=4$, one
magnitude dimmer galaxies ($I\ga28$) were more luminous in the past - indeed,
at the time of reionization (at $z\sim7$ in this simulation) - and by
$z\sim4$ their star formation rate has decreased by a factor of several. In
fact, the galaxy shown in panel ({\it c\/}) has $1.1\times10^9\dim{M}_\sun$
in stars, whereas the galaxy in panel ({\it a\/}), while being more than a
magnitude brighter, has a stellar mass of only $7\times10^8\dim{M}_\sun$.
We, therefore, conclude that galaxies that reionized the universe at 
$z\sim7$ appear at $z\sim4$ within the magnitude range of $28\la I\la30$,
while LBGs that are observed at $z\sim4$ with $23\la I \la25$ are indeed
very young, experiencing their first major episode of star formation (we
refrain from using the word ``burst'', because the star formation history
of a galaxy shown in panel {\it a\/}, while strongly rising by $z\sim4$,
can still be hardly called a ``burst'').

\begin{figure}[t]
\plotonerot{\figdir/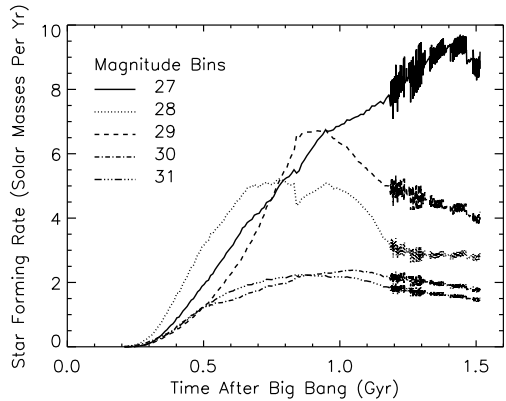}
\caption{\label{fig4}Total star formation rates in galaxies that appear
within a
given magnitude bin at $z=4$
as a function of cosmic time. Each indicated magnitude is the
central value of a bin of width 1.0 magnitudes.}
\end{figure}
Figure \ref{fig4} illustrates this point further. It 
shows total star formation rates for galaxies
that appear within different magnitude bins at $z=4$. In agreement with
individual examples from Fig.\ \ref{fig3}, the star formation rate at
$z=4$ is dominated by our brightest galaxies, but these galaxies contribute
only about 20\% to the star formation at $z\sim7$. At that redshift more
than half of the total star formation rate is contributed by galaxies
that appear within the $28<I<30$ magnitude bin at $z\sim4$.

\begin{figure}[t]
\plotonerot{\figdir/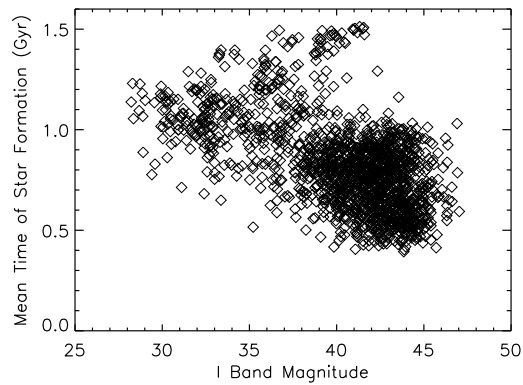}
\caption{\label{fig5}Mean star formation time for all simulated galaxies
as a function of their magnitudes at $z=4$.}
\end{figure}
The large variation in possible star formation histories is further
confirmed by Figure \ref{fig5}, which shows mean mass-weighted star
formation ages for all simulated galaxies as a function of their 
magnitudes at $z=4$,
\[
  t_{SF} = {1\over M_*} \int_0^{t_f} t\,\, {\rm SFR}(t)\, dt,
\]
where $M_*$ is the total stellar mass.
Even for our brightest galaxies the variation is significant,
reaching up to a factor of 3 for the dimmest ones. A remarkable feature
of Fig.\ \ref{fig5} is a large concentration of galaxies with
$40<I<45$ and little star formation after $1\dim{Gyr}$. It is tempting to
identify these galaxies with dwarf spheroidals: they are old 
and would have luminosities of the order of $10^{5 - 6}L_\sun$ at
$z=0$. A small fraction of these galaxies still have active star
formation at $z\sim 4$, which may explain the diverse star formation
histories of dwarf spheroidals in the Local Group (Grebel 1998; Mateo
1998). We leave this as a possible speculation, as this is not the goal of
this paper, however.

Observations at longer wavelengths and/or at fainter magnitudes
would be more sensitive to
older populations of stars. Unfortunately, these observations are
very difficult to make at a redshift of 4, and they will most likely
have to wait for JWST and 30-meter class ground telescopes. But those
instruments are on the horizon, and one might hope to indeed observe
the culprits of reionization within the 10-year time frame.

\begin{figure}[t]
\plotonerot{\figdir/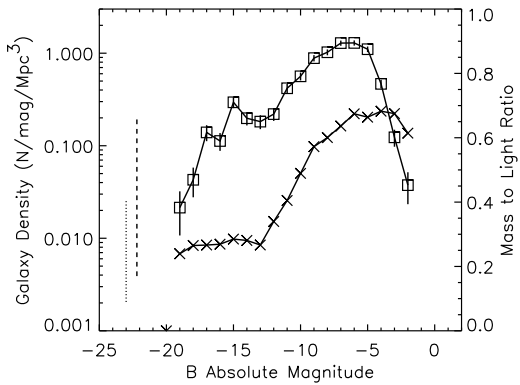}
\caption{\label{fig6}
Mass-to-light ratio for simulated galaxies at $z = 4$.  Squares show the
luminosity function of the simulated galaxies as would be observed
in the K band with $E(B - V) = 0.15$ reddening.  Vertical lines 
through the symbols represent 
1-sigma Poisson errors. Crosses show the stellar mass-to-light ratio
in solar units computed with the same reddening.
The vertical lines on the left show the stellar mass-to-light ranges found by
Dickinson et al.\ (2003) assuming metallicities of $1/5$ solar
(dotted) and solar (dashed).
}
\end{figure}
Whatever the global history of star 
formation we would at least expect that the total stellar 
content of the universe would be monotonic with time.
Thus, we have compared the stellar content of the simulated
galaxies with the global history of total stellar mass as
presented in Dickinson et al.\ (2003). In Figure \ref{fig6} we show
the B band rest frame luminosity function and mass-to-light ratios of 
the simulated galaxies, calculated from the K band magnitudes.
Vertical lines give the comparison with Dickinson et al.\ (2003)
estimates. Our results are in comfortable agreement with these
estimates for all galaxies brighter than dwarf spheroidal candidates.

\begin{figure}[t]
\plotonerot{\figdir/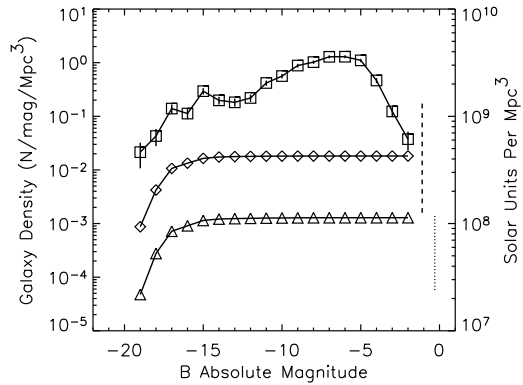}
\caption{\label{fig7}
Cumulative stellar mass and B band luminosity.  Squares show the
luminosity function of the simulated galaxies at $z = 4$ as would be observed
in the K band.  Vertical lines represent 1-sigma Poisson errors.
Triangles show the total stellar mass in simulated galaxies integrated
from the lowest magnitude to the plotted magnitude.  Diamonds trace
the luminosity density of simulated galaxies as would be observed in the
K band with $E(B - V) = 0.15$ reddening.  The dotted line is the
mass density range observed by Dickinson et al.\ (2003).  The dashed
line is their observed luminosity density range obtained by 
integrating over a Schechter fit.
}
\end{figure}
Finally, Figure \ref{fig7} shows cumulative mass and luminosity densities
of simulated galaxies as compared with Dickinson et al.\ (2003) estimates.
Again it appears that the simulated galaxies are in a reasonable agreement
with the observational data. We would like to emphasize that we are
quite comfortable with a factor of 2 agreement: one should remember that
the simulation can only account for about 1/3 of the total light because
of the small box size. In fact, in the imaginary ``super-simulation'',
discussed above, the total stellar mass would be up to a factor of 3 higher
(depending on whether the mass-to-light ratio goes down for brighter galaxies)
and would not agree with Dickinson et al.\ (2003). One could hypothesize
about a source of this potential discrepancy, including
anything from a non-standard IMF to incorrect stellar synthesis models, but
for the purpose of this paper we are content with the level of agreement
we obtain, given substantial limitations of our simulation.

\begin{figure}[t]
\plotonerot{\figdir/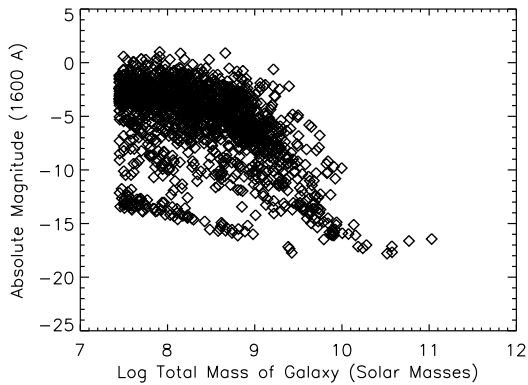}
\caption{\label{fig8}
Scatter plot of absolute magnitude at 1600 A versus total (dark matter +
gas + stellar) mass
of simulated galaxies at redshift 4.  Reddening equivalent to
a color excess of 0.15 has been applied as well as extinction due
to absorption by the Lyman alpha forest.
}
\end{figure}
It would also be interesting to compare the luminosities of Lyman Break
Galaxies and their dark matter masses, because the relation of LBGs and
their dark matter halos is still poorly understood (Somerville, Primack, \& Faber 2001;
Wechsler et al.\ 2001; Somerville 2002; Benson et al.\ 2002a,b).
In Figure \ref{fig8} we show this
relation. As one can see, Lyman Break Galaxies represent a diverse
population. For example, our brightest galaxies with absolute magnitudes 
between -18 and -17 cover about two magnitudes in the total mass: some of
them are the most massive galaxies in the simulations, but some of them
have significantly lower masses and are experiencing a starburst at $z=4$.
There exists a general trend of less massive galaxies being dimmer, but the
scatter around the mean relation is very large.

\section{Discussion}

So, how well does the simulation do? While the computational box of our
simulation is clearly insufficient to have even a single galaxy as luminous
as those of Steidel et al.\
(1999), simulated galaxies do appear as fainter cousins of the
observed ones. They have similar colors, and simulated and observed
galaxies together can be fitted by a single luminosity function. 

They are also faring well in their abilities to reionize the
universe. Star formation histories of simulated galaxies are
diverse, and vary systematically with magnitude. The brightest
galaxies at $z=4$ are indeed quite young, in accord with conclusions of
Shapely et al.\ (2001) and Ferguson et al.\ (2002). But these brightest
galaxies contribute only about 25\% of ionizing photons at $z>6$. 
The bulk of work of reionizing the universe was done
by dimmer galaxies, those that fall within the $28<I<30$ magnitude range at
$z\sim 4$, but which are not necessarily less massive than the bright ones. 

    We expect that detailed simulations will
become increasingly important in sorting out
the various possibilities.  The general picture
that our simulation produces is at least
consistent with observations of Lyman Break
Galaxies, an encouraging message to theorists.

The recent results from the WMAP mission (Kogut et al.\ 2003)
indicate that the optical depth to Thompson scattering is significantly
larger than in our simulation. While this measurement cannot pin down
the redshift of reionization, it suggests, if correct, that
there was a considerable ionizing flux at early times. Our simulation
provides no support for a stellar origin of this ionizing flux. Perhaps,
very massive population III stars, which we have not considered, could be
responsible.

Because,
as we have discussed, the photoionization feedback likely plays a
significant role in the evolution of faint Lyman Break Galaxies, one might
wonder how the WMAP results affect the conclusions of this paper. As has
been shown by several previous studies (Gnedin 2000b and references
herein), the photoionization feedback results in the substantial loss of
the gas mass of objects with circular velocities below about $40\dim{km/s}$.
However, this characteristic circular velocity is essentially redshift 
independent.
So even if the universe was fully ionized throughout its entire history,
the effect of the photoionization feedback at $z=4$ would be essentially at
the same level as in our simulations. Our conclusion about a reasonable
agreement between the simulation and the data is, in fact, independent of
the redshift of reionization as long as it is above about 6.

\acknowledgements

We are grateful to Chuck Steidel for providing us with his custom filter
shapes and for enlightening comments during his visit to CU. This work was
supported by NSF grant AST-0134373. This work was also 
partially supported by National Computational
Science Alliance under grant AST-960015N and utilized 
the SGI/CRAY Origin 2000 array
at the National Center for Supercomputing Applications (NCSA).

\end{document}